\documentclass[pra,twocolumn,showpacs,preprintnumbers,amsmath,amssymb,longbibliography]{revtex4-1}
\usepackage{graphicx,natbib}
\usepackage{dcolumn}
\usepackage{bm}
\usepackage{float}
\usepackage{textcomp} 
\usepackage{color}
\usepackage{colortbl}
\usepackage{hhline}
\usepackage{enumerate}
\definecolor{vortex_grey}{gray}{0.5}
\definecolor{vortex_green}{RGB}{0,200,80}
\definecolor{vortex_blue}{RGB}{0,80,180}

\begin{document}

\title{Einstein--Bose Condensation of Onsager Vortices}

\author{Rahil N. Valani$^{1,2}$, Andrew J. Groszek$^1$, Tapio P. Simula$^1$}
\affiliation{$^1$School of Physics and Astronomy, Monash University, Victoria 3800, Australia}
\affiliation{$^2$Department of Mechanical and Aerospace Engineering, Monash University, Victoria 3800, Australia}

\begin{abstract}
We have studied statistical mechanics of a gas of vortices in two dimensions. We introduce a new observable---a condensate fraction of Onsager vortices---to quantify the emergence of the vortex condensate. The condensation of Onsager vortices is most transparently observed in a single vortex species system and occurs due to a competition between solid body rotation (c.f. vortex lattice) and potential flow (c.f. multiple quantum vortex state). We propose an experiment to observe the condensation transition of the vortices in such a single vortex species system.
\end{abstract}

\maketitle
\section{Introduction}

Perhaps the most astonishing aspect of turbulence is not the complexity of its dynamics but rather that it feeds the emergence of ordered structures out of chaos. The ubiquity of large eddies in two-dimensional fluid flows was also noted by Onsager who suggested that it might be possible to obtain a statistical mechanics description of hydrodynamic turbulence of two-dimensional flows based on discrete collections of point-like vortex particles \cite{Onsager1949a}. In particular, Onsager predicted that turbulent two-dimensional systems could support large scale clustered vortex structures, later coined Onsager vortices, and that such structures would correspond to negative absolute temperature states of the vortex degrees of freedom \cite{Eyink2006a}. Notwithstanding the negative absolute Boltzmann temperature states were observed in nuclear spin systems \cite{Purcell1951a,Oja1997a,Abraham2017a} soon after Onsager's theoretical prediction, and more recently in the motional degrees of freedom of cold atoms confined in optical lattices \cite{Braun2013a}, the negative temperature Onsager vortex states are yet to be uncovered in their original context of two-dimensional (super)fluid turbulence \footnote{The first experimental observation of negative absolute temperature Onsager vortex states have recently been reported in \cite{Gauthier2018a,Johnstone2018a}}.

Kraichnan developed the theory of two-dimensional turbulence further, conjecturing that a scale invariant inverse energy cascade mechanism of incompressible kinetic energy could dynamically lead to the formation of Onsager vortices and even to their condensation and that: \emph{``The phenomenon is analogous to the Einstein--Bose condensation of a finite two-dimensional quantum gas"} \cite{Kraichnan1967a}. In the Kraichnan model the system scale Onsager vortex clusters would emerge due to a termination of the inverse cascade that accumulates energy at ever larger spatial scales. Ultimately, such a process could potentially lead to the condensation of the Onsager vortices, which correspond to the highest accessible energy states of the vortex degrees of freedom \cite{Kraichnan1967a,Kraichnan1975a}. In a neutral system with $N_{\rm tot}$ vortices in total, the condensation of Onsager vortices occurs at a critical negative temperature $T_{\rm EBC} = - \alpha N_{\rm tot} / 4$ \cite{Kraichnan1980a,Viecelli1995a,Simula2014a}, where $\alpha= \rho_s\kappa^2 /4\pi k_{B}=T_{\rm HH} $ is the critical positive temperature for the Hauge--Hemmer pair-collapse transition \cite{Salzberg1963a,Hauge1971a}, which in the case of non-zero vortex core size becomes renormalised to the Berezinskii--Kosterlitz--Thouless (BKT) critical temperature $T_{\rm BKT} = T_{\rm HH}/2$ \cite{Berezinskii1971a,Berezinskii1972a,Kosterlitz1973a}. Here $k_{\rm B}$ is the Boltzmann constant, $\rho_s$ is the (super)fluid density and $\kappa=h/m$ is the circulation quantum with $h$ the Planck's constant and $m$ the particle mass. Inspired by Kraichnan's insight \cite{Kraichnan1967a,Kraichnan1975a}, we refer to the critical temperature of condensation of Onsager vortices with the acronym (EBC), which stands for Einstein--Bose condensation and in the case of zero-core point vortices is also known as supercondensation \cite{Kraichnan1975a}. 

The recent developments of imaging and manipulating compressible superfluids has sparked renewed interest in Onsager's statistical hydrodynamics theory of turbulence. Experiments employing harmonically trapped Bose--Einstein condensates of atoms have ranged from studies of dynamics of vortex dipoles \cite{Neely2010a} or few vortices \cite{Navarro2013a} to three- \cite{Henn2009a} and two-dimensional \cite{Neely2013a,Kwon2014a,Kwon2016a,Seo2016a} quantum turbulence. Moreover, uniform atom traps are becoming increasingly popular \cite{Henderson2009a,Gaunt2013a,Chomaz2015a,Bell2016a,Lee2015a,Gauthier2016a,Mukherjee2017a,Hueck2018a,Johnstone2018a,Gauthier2018a} and will be particularly useful for studies of quantum turbulence. This is partly because well defined trap walls enhance the vortex clustering signal in comparison to harmonically trapped systems \cite{Simula2014a,Groszek2016a,Kwon2014a,Stagg2015a}. In the latter case, strong clustering has not been observed although in both cases Onsager vortices in decaying two-dimensional quantum turbulence has been predicted to emerge via an evaporative heating mechanism of vortices \cite{Simula2014a,Groszek2016a}.

The successes of the recent experimental developments have also spawned novel theoretical investigations \cite{White2012a,Reeves2013a,Billam2014a,Simula2014a,Reeves2015a,Stagg2015a,Groszek2016a,Skaugen2016a,Yu2016a,Skaugen2016b,Salman2016a,Yu2017a,Groszek2017a}. In addition to visual inspection, the presence of Onsager vortices has been associated with indicators such as the vortex dipole moment \cite{Simula2014a,Groszek2016a}, vortex clustering measures \cite{White2012a,Reeves2013a,Skaugen2016a}, or a peak in the power spectral density of incompressible kinetic energy \cite{Yatsuyanagi2005a,Simula2014a,Billam2014a}. However, a measurable that would quantify the degree of condensation of the vortices as opposed to their clustering, has been lacking. Here we use \emph{a vortex--particle duality} to define a condensate fraction that enables quantitative measurements of condensation of Onsager vortices in these two-dimensional systems \cite{Valani2016a}. We have implemented a vortex classification algorithm based on the prescription by Reeves \emph{et al.} \cite{Reeves2013a}, which can be used as a quantitative measure of vortex clustering. Together with the condensate fraction measurable introduced here that uniquely identifies the condensate of Onsager vortices, these two observables enable acquisition of detailed information on clustering and condensation of vortices. We find that the condensate fraction exhibits universal behavior independent of the number of vortices in the bounded circular system. In contrast to condensation, clustering of vortices is present at all negative temperatures in the sense that the total number of vortices belonging to vortex clusters of varying size is greater than zero \cite{Groszek2017a}. Vortex clustering is a precursor to the condensation of Onsager vortices and is reminiscent of the quasi-condensation that precedes the superfluid phase transitions in low-dimensional quantum gas systems \cite{Kagan2000a,Hadzibabic2006a,Clade2009a,Hung2011a}. 

\section{Vortex-Particle duality }

We first consider $N_{\rm tot}$ singly quantised point-like vortices with a hard core of radius $\xi$ and equal numbers of clockwise and counter-clockwise circulations confined in a circular disk of radius $R_\circ$, unless stated otherwise. The pseudo-Hamiltonian describing our system is \cite{Pontin1976a,Simula2014a}:
\begin{eqnarray}
  \label{hami}
  H   =&\alpha k_{\rm B}&\sum_{j}{{s_j^2 \ln(1-r_j^2)} - \alpha k_{\rm B}\sum_{i<j}s_i s_j \ln(r_{ij}^2)} \\
     +&\alpha k_{\rm B}&\sum_{i<j}{s_i s_j \ln(1-2x_i x_j-2y_i y_j+r_i^2 r_j^2)},
\end{eqnarray} 
where $r^2_j = x_j^2+y_j^2$ and $x_j$ and $y_j$ are the dimensionless Cartesian coordinates of the $j$th vortex measured in units of the system radius $R_\circ$ and $s_j=\pm1$ determines the circulation direction of the $j$th vortex. The first, single-vortex, logarithmic term is due to the interaction of each vortex with its own image, the second represents the pairwise two-dimensional Coulomb-like interaction between $i$th and $j$th vortex separated by distance $r_{ij}$ and the last term, due to the circular boundary, represents the interaction of system vortices with the images of all other vortices. 

The dynamics of the point-like vortices are determined by the equations of motion \cite{Onsager1949a}
\begin{equation}
hs_j\frac{\partial {x}_j}{\partial t} = \frac{\partial H}{\partial y_j}\hspace*{5mm}{\rm and}\hspace*{5mm}hs_j\frac{\partial {y}_j}{\partial t} = -\frac{\partial H}{\partial x_j}.
\label{hamdyn}
\end{equation}

To draw a closer correspondence with Hamiltonian mechanics, we may assign for each vortex a canonical coordinate $q_j = R_\circ x_j$ and momentum $p_j = -m_v\omega_0 R_\circ y_j$, where $m_v$ is the vortex mass \cite{Simula2017a} and $\omega_0$ is an angular frequency. Thus the set of vortex coordinates $\{x_j,y_j\}$ in the real space are mapped onto points in the phase space $\{q_j,p_j\}$ spanned by the canonical conjugate variables. In this Hamiltonian description the vortex particles move in one-dimensional real space tracing out orbits in the two-dimensional phase space, which is bounded by the circular wall of radius $R_\circ$. Equation (\ref{hamdyn}) establishes the \emph{vortex--particle duality}---that a vortex in a two-dimensional (2D) fluid may behave as a particle in a one-dimensional (1D) space. 

Motivated by the vortex--particle duality and in contrast to Kraichnan's conjecture, we anticipate the condensation of Onsager vortices to be analogous to the condensation of a finite \emph{one-dimensional} quantum gas. Interestingly, in the 2D fluid picture the vortex condensate corresponds to maximum kinetic energy states of the fluid whereas in the 1D dual picture the condensate corresponds to zero momentum state of the 1D vortex particles.

\section{Ideal vortex gas approximation }
By ignoring the vortex--vortex interactions we obtain an ideal-gas model of vortex particles. A Maclaurin series expansion of the single vortex term in Eq.~(\ref{hami}) with respect to $r_j$ formally yields a one-dimensional harmonic oscillator Hamiltonian
\begin{eqnarray}
  H_0   &=&\alpha k_{\rm B}\sum_{j}{s_j^2 \ln(1-r_j^2)} \\
  &\approx& -\frac{\rho_s\kappa^2}{2\pi} \frac{1}{m_v\omega_0^2R^2_\circ} \sum_{j} \left ( \frac{p_j^2}{2m_v} + \frac{1}{2}m_v\omega_0^2q_j^2\right)\\
  \label{H0}
\end{eqnarray} 
with an inverted energy spectrum with respect to the canonical case. Within the harmonic approximation, a single vortex $v$ of this system will travel along a periodic phase space orbit $\{q_v,p_v\} = \{R_v \cos(\omega_v t), -m_v\omega_0 R_v \sin(\omega_v t)\}$, with orbital angular frequency $\omega_v$ and semi-axis $R_v$.

The Einstein--Brillouin--Keller semiclassical quantisation rule \cite{Keller1958a}
\begin{equation}
\oint p_vdq_v = \left(n+\frac{k}{4}\right)h,
\label{EBK1}
\end{equation}
where $n$ is the principal quantum number and $k$ is the Keller--Maslov index then evaluates to 
\begin{equation}
\int_0^{T} p_vdq_v = \int_0^{\frac{2\pi}{\omega_v}} \omega_v m_v\omega_0 R_v^2\sin^2 (\omega_v t) dt = \pi m_v\omega_0 R_v^2 ,
\label{EBK2}
\end{equation}
where we have integrated over one period, $T=2\pi/\omega_v$, of the vortex orbit. The one-dimensional oscillatory motion has two classical turning points, $k=2$, and therefore the quantisation rule, the combination of Eqs~(\ref{EBK1}) and (\ref{EBK2}), yields the energy spectrum $E_n = (n+\frac{1}{2})\hbar\omega_0 = \frac{1}{2}m_v\omega_0^2 R_v^2$. This implies a minimum semi-axis $\min(R_v) = \xi$ for the vortex trajectories and yields the zero-point energy $E_0 = \frac{1}{2}m_v\omega_0^2 \xi^2$. In correspondence with the Heisenberg uncertainly relation, $\Delta q\Delta p \gtrsim \hbar/2$, the zero-point energy carries the information that the area $A$ of the phase space is quantised in units of $\hbar = m_v\omega_0 \xi^2$. This reflects the fact that it is not possible to localize the position of the vortex inside an area smaller than the vortex core. Consequently, any zero-core point vortex model with  $\xi=0$ violates Heisenberg uncertainty principle and fails to correctly describe the physics of the condensate of Onsager vortices. It is therefore paramount to introduce a non-vanishing vortex core size in order to describe physics of the low entropy states with $T/T_{\rm EBC} <1$.

\section{Interacting vortex gas approximation}
The velocity field induced by the vortices mediates strong vortex-vortex interactions such that the ideal-vortex approximation is strictly only valid for one vortex near the centre of the disk. However, the second term in Eq.~(\ref{hami}) may be approximated as a mean-field potential by integrating out the spatial scales smaller than the intervortex spacing. 

A neutral superfluid that locally rotates at an orbital angular frequency $\Omega$ with $N_v$ vortices of the same sign mimics the rotation of a classical fluid by having an areal vortex density 
\begin{equation}
n_v = \frac{N_v}{\pi R^2}= \frac{m\Omega}{\pi\hbar}.
\label{Feynmanrule}
\end{equation}
Hence, the mean superfluid velocity is $v(r) = \Omega r$, where $r$ is the distance measured from the centre of such a rotating cluster of vortices with radius $R$. In contrast, in a high-winding number vortex with $N_v$ circulation quanta, the superfluid velocity field $v(r)=\frac{\hbar}{m}\frac{N_v}{r}$ is a gradient of a scalar phase function. In general, the velocity field is therefore
\begin{eqnarray}
v(r) &= N_v \frac{\hbar }{ m} \frac{r}{(R^*)^2} ; & r< R^*,\\
v(r) &= N_v\frac{\hbar}{m}\frac{1}{r};  &r> R^*,
\label{velos}
\end{eqnarray}
which is a combination of solid body rotation for $r< R^*$ and potential flow for $r> R^*$, where $R^*$ is the radius of the vortex cluster. The kinetic energy associated with such a flow field may therefore be approximated by a mean-field interaction 
\begin{eqnarray}
H_{\rm int} =& - \alpha k_{\rm B}\sum_{i<j}s_i s_j \ln(r_{ij}^2) \\
\approx & \int_0^{2\pi} \int_0^{R_\circ} \frac{1}{2}\rho_s v^2(r) r dr d\theta\\
=&\pi \rho_s N_v^2\frac{\hbar^2}{m^2}\left( \frac{1}{4} + \ln\left(\frac{R_\circ}{R^*}\right)\right). 
\label{Hmf}
\end{eqnarray}
The final term in the point vortex Hamiltonian, Eq.~(\ref{hami}) describes the remaining interaction with image vortices and yields an energy shift 
\begin{equation}
 - \mu =\alpha k_{\rm B}\sum_{i<j}{s_i s_j \ln(1-2x_i x_j-2y_i y_j+r_i^2 r_j^2)}.
  \label{Hmu}
\end{equation} 
Combining Eqns (\ref{H0}), (\ref{Hmf}), and (\ref{Hmu}), we thus arrive at the effective 1D vortex-particle Hamiltonian, 
\begin{eqnarray}
H_{\rm eff} &=& H_0 + H_{\rm int} - \mu  \\
        &\approx & -\frac{\rho_s\kappa^2}{2\pi} \bigg[\frac{1}{m_v\omega_0^2R^2_\circ} \sum_{j} \left ( \frac{p_j^2}{2m_v}+\frac{1}{2}m_v\omega_0^2q_j^2\right)\\
 &-& \frac{1}{2}N^2_v\left( \frac{1}{4} + \ln\left(\frac{R_\circ}{R^*}\right)\right)  \bigg] -\mu
\label{heff}
\end{eqnarray} 
that describes a system of one-dimensional strongly interacting harmonic oscillators. It may be worth pointing out that the two forms of Eq.~(\ref{Hmf}) have quite different interpretations. The first line is a long-range interaction of the vortices in 2D, whereas the last line is a strong contact interaction between 1D vortex particles with a coupling strength that is running with the energy scale set by the radius $R^*$ of the cluster. In the Tonks-Girardeau-like limit of $R^*\to0$ the effective coupling constant $g_1\propto \ln\left(\frac{R_\circ}{R^*}\right)\to \infty$ and Eq.~(\ref{heff}) reduces to a semi-classical version of the Lieb-Liniger model \cite{Giamarchi2003a}.

\begin{figure}[t]
\centering
\includegraphics[width=0.8\columnwidth]{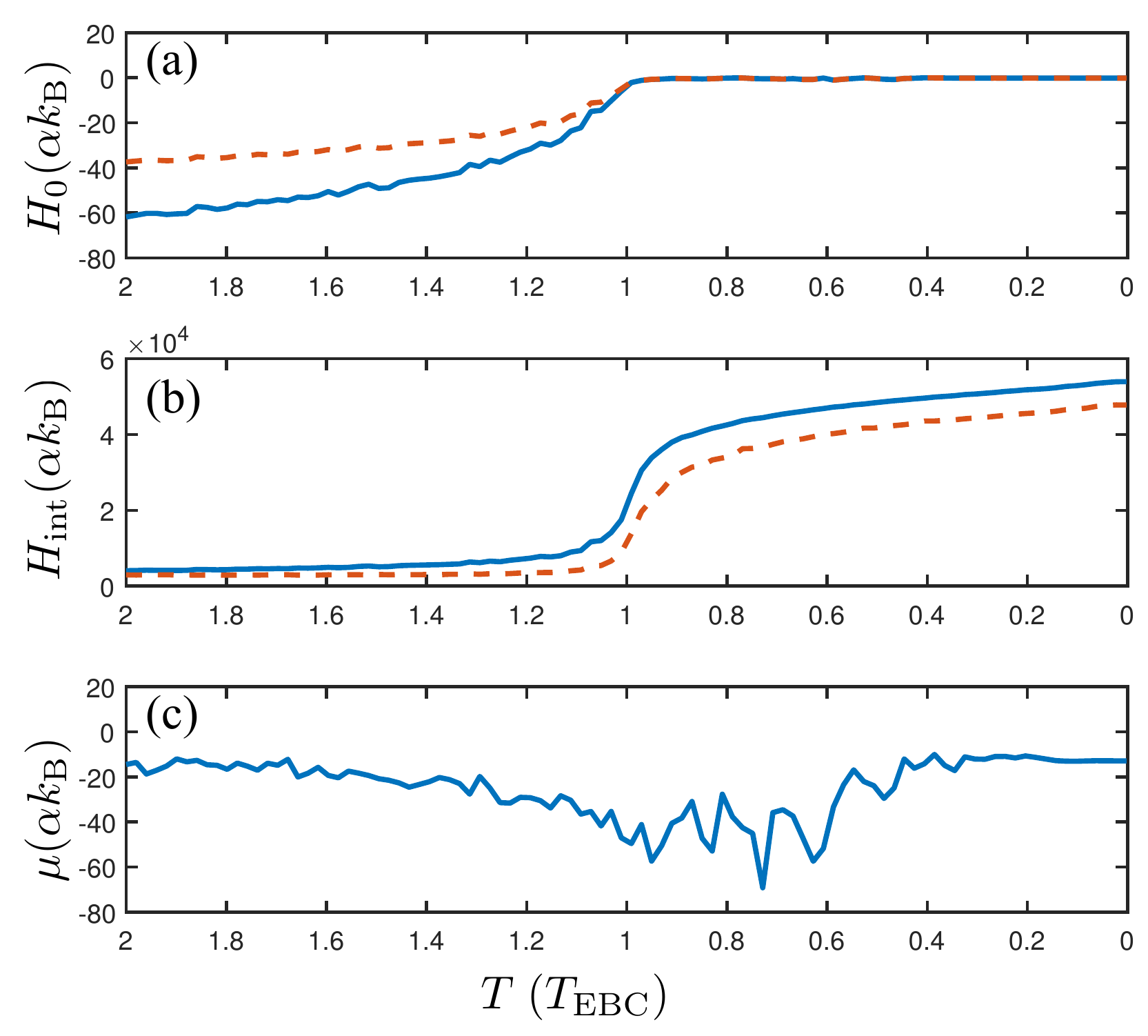}
\caption{
   The three energy terms (a)-(c) of Eq.~(\ref{hami}), solid lines, and the respective approximations Eq.~(\ref{H0}) and Eq.~(\ref{Hmf}), dashed lines, as functions of reduced temperature for $N_v=100$ single species, $s=+1$, vortices. 
}\label{hamterms}
\end{figure}

Figure \ref{hamterms} shows the independent contributions of the three terms in the Hamiltonian, Eq.~(\ref{hami}), for a system of 100 like-signed vortices as functions of reduced temperature. The details of this calculation are described in Sec.~\ref{PolarisedMC}. For comparison, the energy contributions due to the harmonic oscillator and mean-field approximations, Eq.~(\ref{H0}) and Eq.~(\ref{Hmf}), respectively, are shown by dashed lines. The harmonic oscillator approximation, Eq.~(\ref{H0}), is better at lower reduced temperatures because the vortices clump close to the centre of the disk. However, since the mean-field term, Eq.~(\ref{Hmf}), is proportional to $N^2_v$, it is overwhelmingly larger than the single vortex terms, which are proportional to $N_v$. These results establish that the mean-field Hamiltonian Eq.~(\ref{heff}) is a reasonable approximation for Eq.~(\ref{hami}).

\section{Fraction of condensed vortices} 
On the basis of the vortex-particle duality, we anticipate condensation of Onsager vortices when the phase space density $n_v \lambda_v \gtrsim 1$. Here $n_v$ is the one-dimensional mean vortex density and 
\begin{equation}
\lambda_v = \frac{h}{\langle p\rangle} \sim \frac{2\pi\xi^2}{\langle R_v\rangle}
\end{equation}
is the thermal vortex de-Broglie wavelength, which in the vortex dual is inversely proportional to the size of an average temperature-dependent vortex orbit in the phase space. For $N^*$ vortices confined within length $2R^*$ the condensation criterion becomes $\pi N^* \xi^2 / \langle R_v\rangle R^*  \sim 1$, which shows that condensation is expected when the vortices concentrate into a phase-space cluster with size of the order of $\xi\sqrt{N^*}$. 

These considerations lead us to define the fraction of condensed vortices as the ratio, $N_0/N$, of $N_0$ vortices of a given sign in a single many-vortex cluster to the total number of vortices $N$ of that same sign in the system. The highest density of vortices is found within clusters and by denoting $N^*$ to be the number of vortices in the largest cluster, which will be the first to condense, and $A^0=N^*\hbar=N^*m_v\omega_0\xi^2$ and $A^*=N^*m_v\omega_0\langle r_{\rm nn} \rangle^2$ to be, respectively, the minimum possible phase space area occupied by the $N^*$ vortices and the phase space area actually covered by them, we obtain  
\begin{equation}
\frac{N_0}{N} =\frac{N^*}{N}\frac{A^0}{A^*} = \frac{N^*}{N}\frac{\xi^2}{\langle r_{\rm nn} \rangle^2}. 
\label{EBCfraction}
\end{equation}
Thus the condensate fraction is the product of the largest cluster fraction $N^*/N$ and the square of the ratio of single vortex core radius $\xi$ to the mean radius $\langle r_{\rm nn}\rangle$ of the effective area occupied by a vortex within the cluster, where $r_{\rm nn}$ is one half of the distance between the centres of nearest neighbour vortices in such a cluster. Although for single vortex species systems $N^*/N=1$, in general, the system contains both vortices and antivortices and to measure $N^*<N$ in such systems, clusters of like-signed vortices must first be identified by a vortex classification algorithm.

\section{Vortex Classification Algorithm}

\begin{figure}
\centering
\includegraphics[width=0.9\columnwidth]{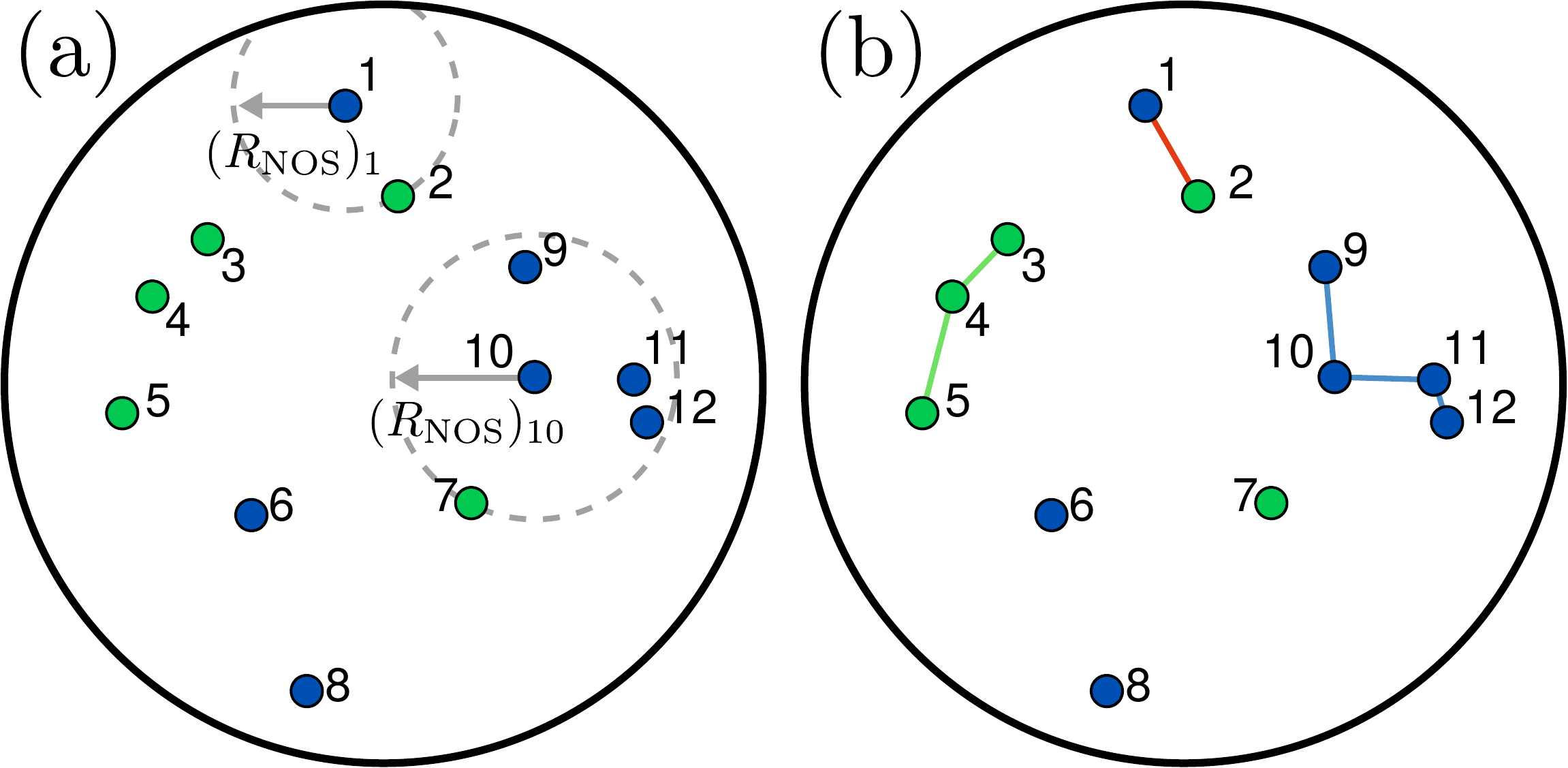}
\caption{A configuration of twelve point vortices, (a) before and (b) after the classification algorithm has been applied. Vortices are drawn in blue, while antivortices are drawn in green. In panel (a), dashed circles are drawn centered on $v_1$ and $v_{10}$, denoting the respective distances $(R_{\rm NOS})_j$ to the nearest opposite signed vortex. Because $v_2$ is the closest vortex to $v_1$ and is of opposite sign, $v_2$ is labelled as a dipole candidate for $v_1$. Vortex $v_{10}$, on the other hand, is closer to $v_9$, $v_{11}$ and $v_{12}$ than it is to $v_7$; hence, these three vortices become cluster candidates for $v_{10}$. The lines joining clustered vortices in (b) are drawn using a minimum spanning tree algorithm, which is applied once all vortices have been labelled into the sets of clusters, dipoles or free vortices.}
\label{fig:cluster_example}
\end{figure}

To quantitatively study clustering and condensation of vortices we have implemented a vortex classification algorithm based on the prescription by Reeves \emph{et al.} \cite{Reeves2013a}. We assign each vortex in a given configuration of $N$ vortices a unique and arbitrarily chosen label from the set $\lbrace v_1, \, v_2, \, \ldots, \, v_N \rbrace$. The vortex configuration is then described by a corresponding set of positions $\lbrace z_1, \, z_2, \, \ldots, \, z_N \rbrace$ (in two-dimensional complex co-ordinates, where $z_j = x_j + i y_j$) and circulation signs $\lbrace s_1,\, s_2,\, \ldots,\, s_N \rbrace$, which here take the value $s_j = \pm 1$, denoting clockwise or anti-clockwise circulation. The algorithm does not prioritise any vortex and yields the same classification outcome regardless of the choice of vortex labelling. Figure \ref{fig:cluster_example} shows an example configuration of twelve judiciously numbered point vortices. The vortex classification algorithm is outlined below.

\begin{figure}[b]
\centering
\includegraphics[width=\columnwidth]{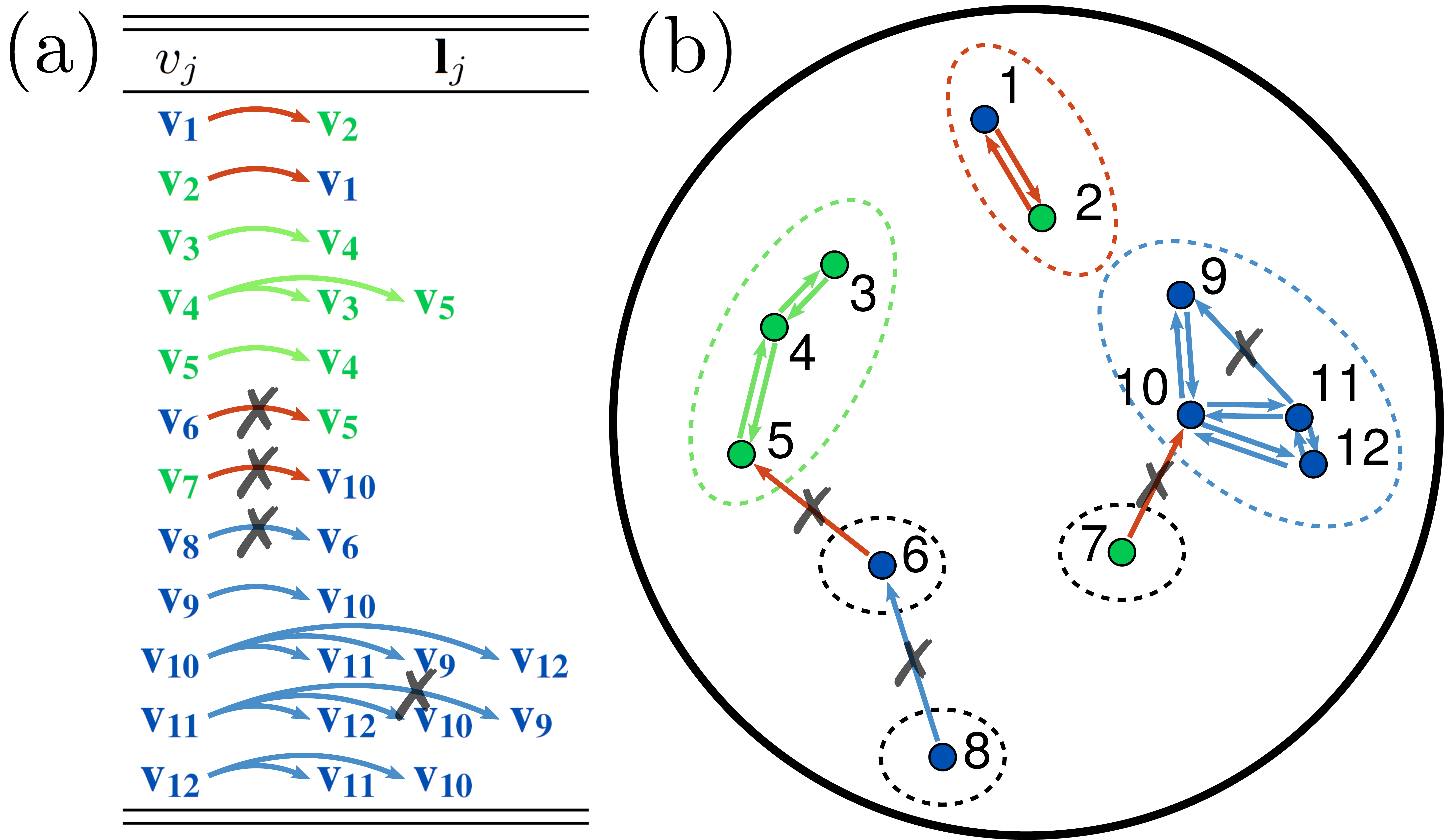}
\caption{The process of identifying mutual neighbours, shown equivalently as (a) a table of candidate lists $\textbf{l}_j$ taken from Table \ref{tab:labels}, and (b) drawn directly onto the example vortex configuration from Fig.~\ref{fig:cluster_example}. An arrow is drawn from each vortex $v_j$ to all the members of its candidate list $\textbf{l}_j$. Only when arrows point in both directions between $v_j$ and $v_k$ are they defined to be mutual neighbours. All arrows that are one-directional have been crossed out in both panels. In panel (b), dashed circles are drawn around clusters (blue/green for positive/negative), dipoles (red) and free vortices (black).}
\label{fig:mutual_neighbour_schematic}
\end{figure}
\subsection{Step 1: Find dipole and cluster candidates}

For each vortex $v_j$, we locate the nearest opposite sign (NOS) vortex and label it as $(v_{\rm NOS})_j$ [\textit{i.e.}~the nearest vortex which satisfies $s_j (s_{\rm NOS})_j < 0$]. We define the distance to this vortex to be $(R_{\rm NOS})_j \equiv \left| z_j - (z_{\rm NOS})_j \right|$. We then check to see if any other vortices (which are same-sign, by necessity) fall within the disk of radius $R_{\rm NOS}$ centered at vortex $v_j$.
\begin{enumerate}[(i)]
\item \textit{Dipoles:} If not, then $(v_{\rm NOS})_j$ is labelled as a dipole candidate for $v_j$ [\textit{e.g.}~in Fig.~\ref{fig:cluster_example}(a), $v_2$ is labelled as a dipole candidate for $v_1$].
\item \textit{Clusters:} If there are $n_j \geq 1$ vortices which are nearer to $v_j$ than $(v_{\rm NOS})_j$, then these are labelled as cluster candidates for $v_j$ [\textit{e.g.}~vortex $v_{10}$ in Fig.~\ref{fig:cluster_example}(a), for which $v_9$, $v_{11}$ and $v_{12}$ are cluster candidates].
\end{enumerate}
Each vortex $v_j$ now has a corresponding set of candidate vortex labels, which we denote by $\textbf{l}_j$. For case (i), $\textbf{l}_j$ consists of a single opposite sign vortex, which is a dipole candidate. For case (ii), $\textbf{l}_j$ is a list of $n_j$ same-sign cluster candidates.

Table \ref{tab:labels} below displays the lists $\textbf{l}_j$ that are constructed in Step 1 of the algorithm when it is applied to the configuration shown in Fig.~\ref{fig:cluster_example}.

\begin{table}[b]

\caption{Collation of $\textbf{l}_j$ lists for the configuration shown in Fig.~\ref{fig:cluster_example} after dipole and cluster candidates have been identified. Each row corresponds to a particular vortex $v_j$ (leftmost column) and the list of all other vortices, ordered from left to right in increasing distance from $v_j$. Vortices/antivortices $v_k$ are denoted with blue/green font if $|z_j - z_k| \leq (R_{\rm NOS})_j$. All vortices for which $|z_j - z_k| > (R_{\rm NOS})_j$ are colored in gray, as these cannot be dipole or cluster candidates. The lists $\textbf{l}_j$ consist of either a single opposite sign vortex (\textit{e.g.}~row 1, corresponding to vortex $v_1$, which has $\textbf{l}_1=\lbrace v_2 \rbrace$), or a set of $\geq1$ same-sign vortices (\textit{e.g.}~row 4, corresponding to vortex $v_4$, for which $\textbf{l}_4=\lbrace v_3, v_5 \rbrace$).}

\label{tab:labels}
\begin{center}

\begin{tabular}{r@{\hskip 20pt}lllllllllll}
\hhline{============}
$v_j$ & \multicolumn{11}{c}{$\textbf{l}_j$}  \\
\hline
{\color{vortex_blue}$\textbf{v}_\textbf{1}$}     & {\color{vortex_green}$\textbf{v}_\textbf{2}$}  & {\color{vortex_grey}$v_3$}  & {\color{vortex_grey}$v_9$}  & {\color{vortex_grey}$v_4$}  & {\color{vortex_grey}$v_{10}$} & {\color{vortex_grey}$v_5$}  & {\color{vortex_grey}$v_{11}$} & {\color{vortex_grey}$v_7$}  & {\color{vortex_grey}$v_6$}  & {\color{vortex_grey}$v_{12}$} & {\color{vortex_grey}$v_8$}  \\

{\color{vortex_green}$\textbf{v}_\textbf{2}$}     & {\color{vortex_blue}$\textbf{v}_\textbf{1}$}  & {\color{vortex_grey}$v_9$}  & {\color{vortex_grey}$v_3$}  & {\color{vortex_grey}$v_{10}$} & {\color{vortex_grey}$v_4$}  & {\color{vortex_grey}$v_{11}$} & {\color{vortex_grey}$v_7$}  & {\color{vortex_grey}$v_{12}$} & {\color{vortex_grey}$v_5$}  & {\color{vortex_grey}$v_6$}  & {\color{vortex_grey}$v_8$}  \\

{\color{vortex_green}$\textbf{v}_\textbf{3}$}     & {\color{vortex_green}$\textbf{v}_\textbf{4}$}  & {\color{vortex_blue}$\textbf{v}_\textbf{1}$}  & {\color{vortex_grey}$v_2$}  & {\color{vortex_grey}$v_5$}  & {\color{vortex_grey}$v_6$}  & {\color{vortex_grey}$v_9$}  & {\color{vortex_grey}$v_{10}$} & {\color{vortex_grey}$v_7$}  & {\color{vortex_grey}$v_{11}$} & {\color{vortex_grey}$v_8$}  & {\color{vortex_grey}$v_{12}$} \\

{\color{vortex_green}$\textbf{v}_\textbf{4}$}     & {\color{vortex_green}$\textbf{v}_\textbf{3}$}  & {\color{vortex_green}$\textbf{v}_\textbf{5}$}  & {\color{vortex_blue}$\textbf{v}_\textbf{6}$}  & {\color{vortex_grey}$v_2$}  & {\color{vortex_grey}$v_1$}  & {\color{vortex_grey}$v_9$}  & {\color{vortex_grey}$v_7$}  & {\color{vortex_grey}$v_{10}$} & {\color{vortex_grey}$v_8$}  & {\color{vortex_grey}$v_{11}$} & {\color{vortex_grey}$v_{12}$} \\

{\color{vortex_green}$\textbf{v}_\textbf{5}$}     & {\color{vortex_green}$\textbf{v}_\textbf{4}$}  & {\color{vortex_blue}$\textbf{v}_\textbf{6}$}  & \color{vortex_grey}$v_3$  & \color{vortex_grey}$v_8$ & \color{vortex_grey}$v_2$ & \color{vortex_grey}$v_7$  & \color{vortex_grey}$v_1$ & \color{vortex_grey}$v_{10}$ & \color{vortex_grey}$v_9$  & \color{vortex_grey}$v_{11}$ & \color{vortex_grey}$v_{12}$ \\

{\color{vortex_blue}$\textbf{v}_\textbf{6}$}     & {\color{vortex_green}$\textbf{v}_\textbf{5}$}  & \color{vortex_grey}$v_8$  & \color{vortex_grey}$v_7$  & {\color{vortex_grey}$v_4$}  & {\color{vortex_grey}$v_3$}  & \color{vortex_grey}$v_{10}$ & {\color{vortex_grey}$v_2$}  & \color{vortex_grey}$v_9$  & \color{vortex_grey}$v_{12}$ & \color{vortex_grey}$v_{11}$ & {\color{vortex_grey}$v_1$}  \\

{\color{vortex_green}$\textbf{v}_\textbf{7}$}     & {\color{vortex_blue}$\textbf{v}_\textbf{10}$} & \color{vortex_grey}$v_{12}$ & \color{vortex_grey}$v_{11}$ & \color{vortex_grey}$v_6$  & \color{vortex_grey}$v_9$  & \color{vortex_grey}$v_8$  & \color{vortex_grey}$v_2$  & \color{vortex_grey}$v_{5}$  & \color{vortex_grey}$v_3$  & {\color{vortex_grey}$v_4$}  & {\color{vortex_grey}$v_1$} \\

{\color{vortex_blue}$\textbf{v}_\textbf{8}$}     & \color{vortex_blue}$\textbf{v}_\textbf{6}$  & {\color{vortex_green}$\textbf{v}_\textbf{7}$}  & \color{vortex_grey}$v_5$  & \color{vortex_grey}$v_{10}$ & {\color{vortex_grey}$v_4$}  & \color{vortex_grey}$v_{12}$ & \color{vortex_grey}$v_{11}$ & \color{vortex_grey}$v_3$  & \color{vortex_grey}$v_9$  & {\color{vortex_grey}$v_2$}  & {\color{vortex_grey}$v_1$}  \\

{\color{vortex_blue}$\textbf{v}_\textbf{9}$}     & \color{vortex_blue}$\textbf{v}_\textbf{10}$ & {\color{vortex_green}$\textbf{v}_\textbf{2}$}  & \color{vortex_grey}$v_{11}$ & \color{vortex_grey}$v_{12}$ & \color{vortex_grey}$v_7$  & {\color{vortex_grey}$v_1$}  & {\color{vortex_grey}$v_3$}  & {\color{vortex_grey}$v_4$}  & \color{vortex_grey}$v_6$  & \color{vortex_grey}$v_5$  & \color{vortex_grey}$v_8$  \\

{\color{vortex_blue}$\textbf{v}_\textbf{10}$}    & \color{vortex_blue}$\textbf{v}_\textbf{11}$ & \color{vortex_blue}$\textbf{v}_\textbf{9}$  & \color{vortex_blue}$\textbf{v}_\textbf{12}$ & {\color{vortex_green}$\textbf{v}_\textbf{7}$}  & {\color{vortex_grey}$v_2$}  & \color{vortex_grey}$v_6$  & {\color{vortex_grey}$v_1$}  & {\color{vortex_grey}$v_3$}  & \color{vortex_grey}$v_8$  & {\color{vortex_grey}$v_4$}  & \color{vortex_grey}$v_5$  \\

{\color{vortex_blue}$\textbf{v}_\textbf{11}$}    & \color{vortex_blue}$\textbf{v}_\textbf{12}$ & \color{vortex_blue}$\textbf{v}_\textbf{10}$ & \color{vortex_blue}$\textbf{v}_\textbf{9}$  & {\color{vortex_green}$\textbf{v}_\textbf{7}$}  & {\color{vortex_grey}$v_2$}  & {\color{vortex_grey}$v_1$}  & \color{vortex_grey}$v_6$  & {\color{vortex_grey}$v_3$}  & \color{vortex_grey}$v_8$  & {\color{vortex_grey}$v_4$}  & \color{vortex_grey}$v_5$  \\

{\color{vortex_blue}$\textbf{v}_\textbf{12}$}    & \color{vortex_blue}$\textbf{v}_\textbf{11}$ & \color{vortex_blue}$\textbf{v}_\textbf{10}$ & {\color{vortex_green}$\textbf{v}_\textbf{7}$}  & \color{vortex_grey}$v_9$  & {\color{vortex_grey}$v_2$}  & \color{vortex_grey}$v_6$  & \color{vortex_grey}$v_8$  & \color{vortex_grey}$v_1$  & {\color{vortex_grey}$v_3$}  & {\color{vortex_grey}$v_4$}  & \color{vortex_grey}$v_5$  \\

\hhline{============}
\end{tabular}
\end{center}
\end{table}

\subsection{Step 2: Find mutually agreeing candidates}

In the second step of the algorithm, the lists $\textbf{l}_j$ are checked sequentially for mutual members. This process is shown schematically in Fig.~\ref{fig:mutual_neighbour_schematic} for the example configuration shown in Fig.~\ref{fig:cluster_example} and Table \ref{tab:labels}.
\begin{enumerate}[(i)]

\item \textit{Dipoles:} If a list $\textbf{l}_j$ consists of a single dipole candidate $v_k$, then the list $\textbf{l}_k$ is checked to see if it contains (only) the vortex $v_j$. If so, then the two vortices are mutual nearest neighbours of opposite sign, and are classified as a dipole (\textit{e.g.}~vortices $v_1$ and $v_2$ in Fig.~\ref{fig:mutual_neighbour_schematic}). If not, then the vortices are left unclassified (\textit{e.g.}~vortices $v_6$ and $v_5$ in Fig.~\ref{fig:mutual_neighbour_schematic}).

\item \textit{Clusters:} If a list $\textbf{l}_j$ consists of a set of cluster candidate vortices $\lbrace v_k \rbrace$, then the lists $\lbrace \textbf{l}_k \rbrace$ are all checked to see if they contain the vortex $v_j$. For each list $\textbf{l}_k$ that does contain $v_j$, the two vortices $v_j$ and $v_k$ are labelled as belonging to the same cluster (\textit{e.g.}~in Fig.~\ref{fig:mutual_neighbour_schematic}, vortex $v_4$ ``checks" both $\textbf{l}_3$ and $\textbf{l}_5$ to see if it is a member of either. It is found to be a member of both, so all three vortices are placed in a single cluster). For each list $\textbf{l}_k$ that does not contain $v_j$, neither vortex label is updated (\textit{e.g.}~vortex $v_7$ and $v_{10}$ in Fig.~\ref{fig:mutual_neighbour_schematic}). Note that not all members of a single cluster have to be mutual candidates of one another. In the example shown  in Fig.~\ref{fig:mutual_neighbour_schematic}, $v_9$ is only a mutual neighbour with $v_{10}$, but is still placed in the same cluster as $v_{11}$ and $v_{12}$. As the algorithm proceeds, vortices may be assigned to existing clusters, or previously classified clusters may become merged.

\end{enumerate}

Any vortices left unclassified after this process are classified as free vortices, as they have no mutual dipole or cluster neighbours (\textit{e.g.}~vortex $v_6$ in Fig.~\ref{fig:mutual_neighbour_schematic}).

In Fig.~\ref{fig:mutual_neighbour_schematic}(b), any two vortices that are connected by a two-directional link are part of the same cluster or dipole, while any vortex that has no two-directional links is a free vortex.

To reduce computation, the checking of mutual candidates can be restricted such that it is only initiated for $v_j$ and $v_k$ if $j > k$. Alternatively, once a pair of vortices has been checked, then $v_j$ could be removed from $\textbf{l}_k$ and vice versa.

\begin{figure*}[!t]
\centering
\includegraphics[width=2\columnwidth]{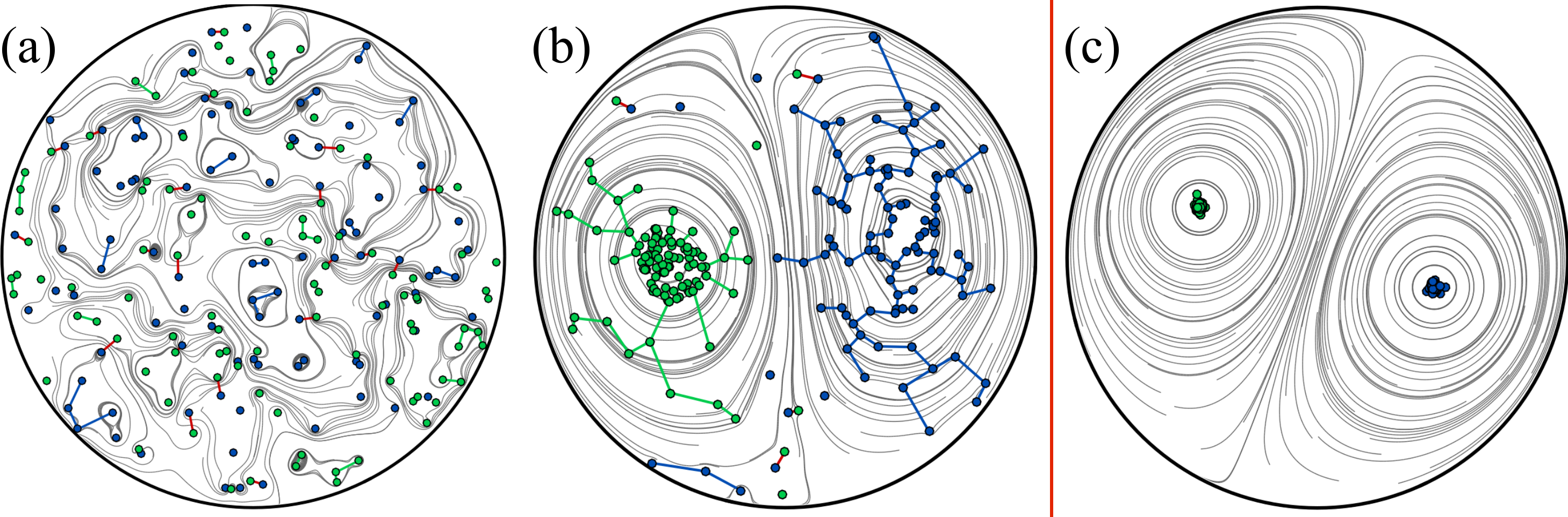}
\caption{   Representative neutral vortex configurations (a)-(c) at respective temperatures $T/T_{\rm EBC}= 10^6,1.022,$ and $0.778$, with $T_{\rm EBC}=-0.25\alpha N_{\rm tot}$ and $N_{\rm tot}=200$. Vortices in vortex and antivortex clusters are connected by blue and green lines, respectively, vortex and an antivortex in vortex dipoles are connected by red lines and free vortices are marked by isolated filled circles. The streamlines illustrate the velocity field generated by the collection of vortices. The red vertical line indicates the location of the condensation transition point $T_{\rm EBC}$ between (b) and (c).}
\label{fig1}
\end{figure*}

\section{Two-species Monte Carlo results}
To study the thermodynamics of the condensation of Onsager vortices, we have performed Monte Carlo calculations using a Metropolis algorithm to find the equilibrium vortex configurations as functions of temperature for systems with $10,20,50,100,200,300$ and $400$ vortices \cite{Viecelli1995a,Simula2014a}. The Monte Carlo calculations, and the conclusions drawn from the results, are obtained using canonical ensemble with hard core vortex core regularization. A hard core diameter of $2\xi=0.001\;R_\circ$ was imposed on each vortex in the results presented. The Monte Carlo samplings were performed for temperature in the range $T\in(-\infty,-0)$ with $10^6$ microstates at each temperature after initial burn in of $10^6$ steps. Out of the $10^6$ microstates, $1000$ uniformly spaced configurations were recorded and used for vortex classification analysis.

\begin{figure}[!hb]
\centering
\includegraphics[width=0.9\columnwidth]{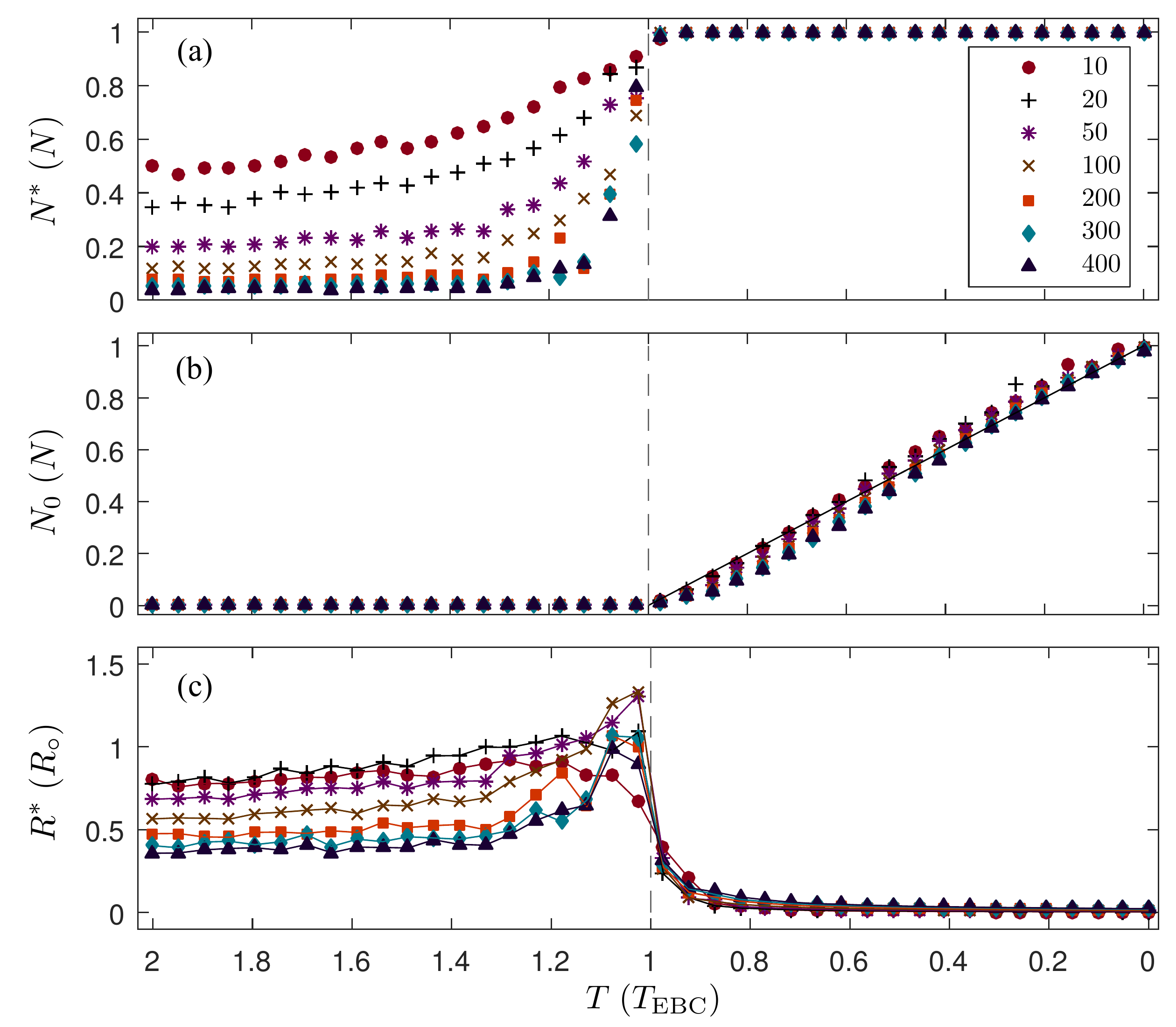}
\caption{  
 Fraction of vortices in the largest cluster (a), the condensate fraction, Eq.~(\ref{EBCfraction}), (b) and the mean cluster radius $R^* = 4 {\rm Std}(r_j)$, where ${\bf r}_j$ is the vector of positions of the vortices in the largest cluster (c) as functions of scaled temperature. Data is shown for systems with different vortex numbers as indicated in the legend. The function $1-T/T_{\rm EBC}$, where $T_{\rm EBC}=-0.25\alpha N_{\rm tot}$, is shown as a solid line in (b) for $|T|<|T_{\rm EBC}|$ and the vertical dashed line marks the critical point. All quantities are ensemble averaged.
}
\label{fig2}
\end{figure}

\begin{figure}[!t]
\centering
\includegraphics[width=0.8\columnwidth]{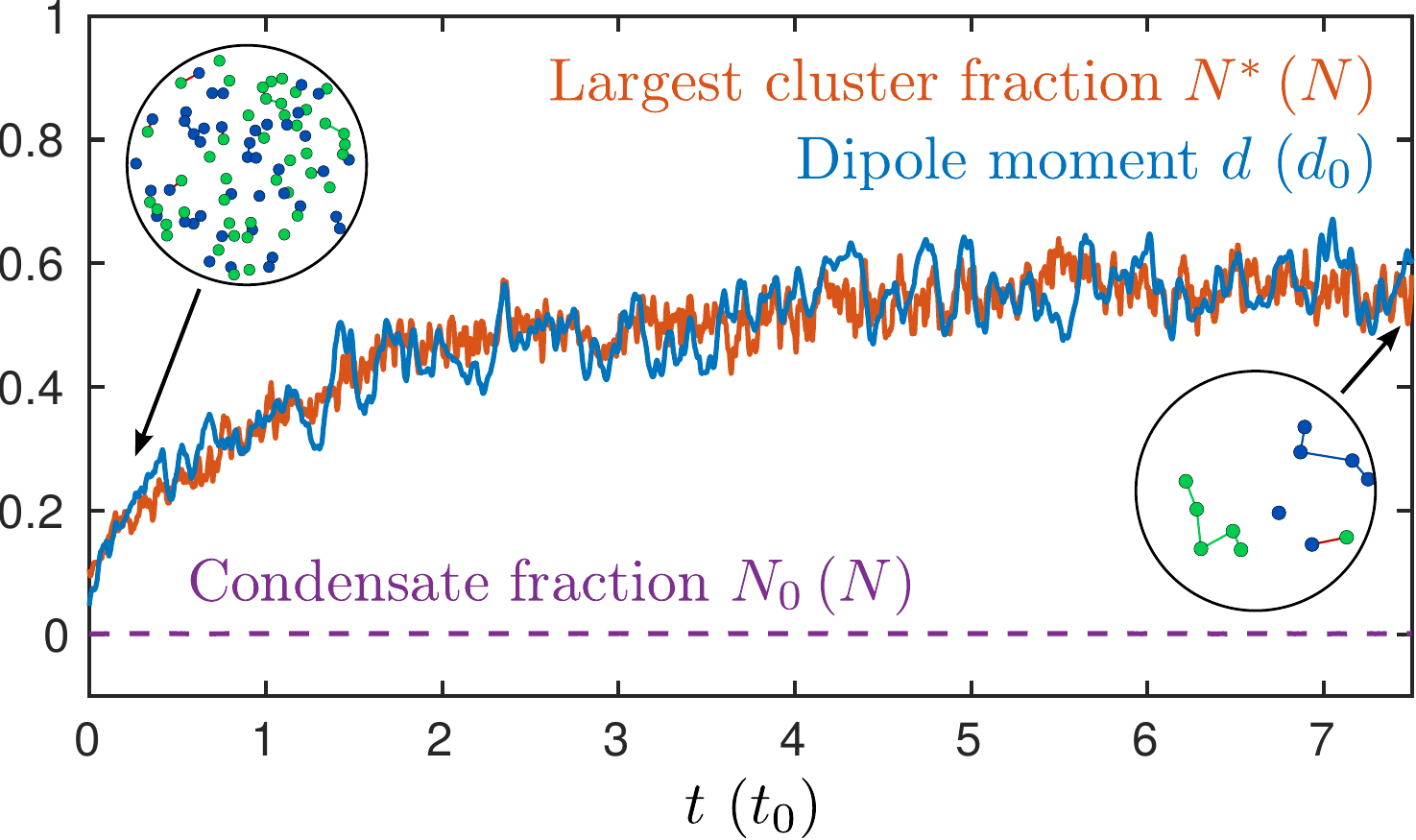}
\caption{
   Dipole moment of the vortex configuration (solid blue) expressed in units of $d_0=\kappa R_\circ N$, largest cluster fraction (solid orange) and condensate fraction (dashed purple) as functions of time calculated from a dynamical mean-field simulation. The unit of time is $t_0= \hbar/\mu_{\rm GPE}$, where $\mu_{\rm GPE}$ is the chemical potential of the Gross--Pitaevskii equation. The initial vortex number is 100 and as the system evaporatively heats up, the vortex number decays to a value of 12 at the end of the simulation, see Ref.~\cite{Groszek2016a} for details. 
}\label{fig3}
\end{figure}


Figure~\ref{fig1} shows typical vortex configurations of disordered and strongly clustered neutral vortex states of $N_{\rm tot} = 200$ vortices obtained from the Monte Carlo calculations at different temperatures. The same sign clusters, dipoles and free vortices are identified using the vortex classification algorithm and the velocity field stream lines are included to visualise the superflow around the vortices. Figure \ref{fig1}~(a)  shows a vortex configuration at a high negative temperature $T = 10^6\;T_{\rm EBC}$ revealing a fairly disordered configuration of vortices with an abundance of vortex dipoles and small clusters. Figure \ref{fig1}~(b) shows a vortex configuration at $T=1.022\;T_{\rm EBC}$ close to the critical temperature. In Fig.~\ref{fig1}~(b) nearly all the vortices have already clustered into two large Onsager vortices although the condensate fraction remains zero. The prominent dipolar shape of the streamlines in (b) and (c) is observable before the critical point $T=T_{\rm EBC}$, indicated by the red vertical line, see also Supplemental Figure~S1(b) of Ref.~\cite{Simula2014a}. Figure \ref{fig1}~(c) shows Onsager vortices at temperature $T= 0.778 \;T_{\rm EBC}$ where the system has a condensate fraction of $N_0/N\approx 0.1$. The qualitative similarity between the stream lines in Figs \ref{fig1}~(b) and (c) is striking despite the states lying on different sides of the transition.

Figure \ref{fig2} shows (a) the largest cluster fraction,  (b) the condensate fraction, and (c) the mean radius of the largest cluster in the system as functions of temperature in units of the critical temperature $T_{\rm EBC} = -0.25 \alpha N_{\rm tot}$. The largest cluster fraction Fig.~\ref{fig2}(a) is strongly dependent on the total number of vortices in the system. In contrast, the condensate fraction, shown in Fig.~\ref{fig2}(b), remains zero at all temperatures $|T|> |T_{\rm EBC}|$ and thereafter increases as the absolute negative zero is approached. Figure~ \ref{fig2}(c) shows the mean radii of the largest vortex clusters as functions of temperature. As the critical temperature is approached from the disordered side, the largest cluster tends to grow in size as ever more vortices are joining the cluster. In the condensed phase the cluster rapidly shrinks as the phase-space density, and hence the condensate fraction, increases. Importantly, the condensate fraction shows universality in the sense that it is consistent with data collapsing onto a single curve, indicating the condensate fraction becoming a vortex number independent quantity in the large vortex number limit. 

With the ability to quantify the condensation of Onsager vortices, we have revisited the dynamical mean-field simulations of Ref.~\cite{Groszek2016a}. Figure \ref{fig3} shows a typical result revealing that in this neutral vortex system, the largest cluster fraction and vortex dipole moment are practically equivalent observables. However, although the system is continually evaporatively heated, the condensate fraction remains zero for all times. The initial vortex number in this simulation is 100 and it decays to the final value of 12. Comparing the largest cluster fraction in Fig.~\ref{fig3} with Fig.~\ref{fig2}(a) shows that this system is initially at temperature $|T|\gg \; |T_{\rm EBC}|$ and evaporatively heats reaching a final temperature of $|T|\gtrsim \;|T_{\rm EBC}|$. Quantitatively, the temperature of the vortex system can be found using the vortex thermometry based on the fraction of clustered vortices in the system \cite{Groszek2017a}. However, once the system becomes fully clustered, the evaporative heating mechanism switches off \cite{Simula2014a} and the condensation is unable to proceed.

\section{One-species Monte Carlo results}\label{PolarisedMC}
\begin{figure*}[!t]
\centering
\includegraphics[width=2\columnwidth]{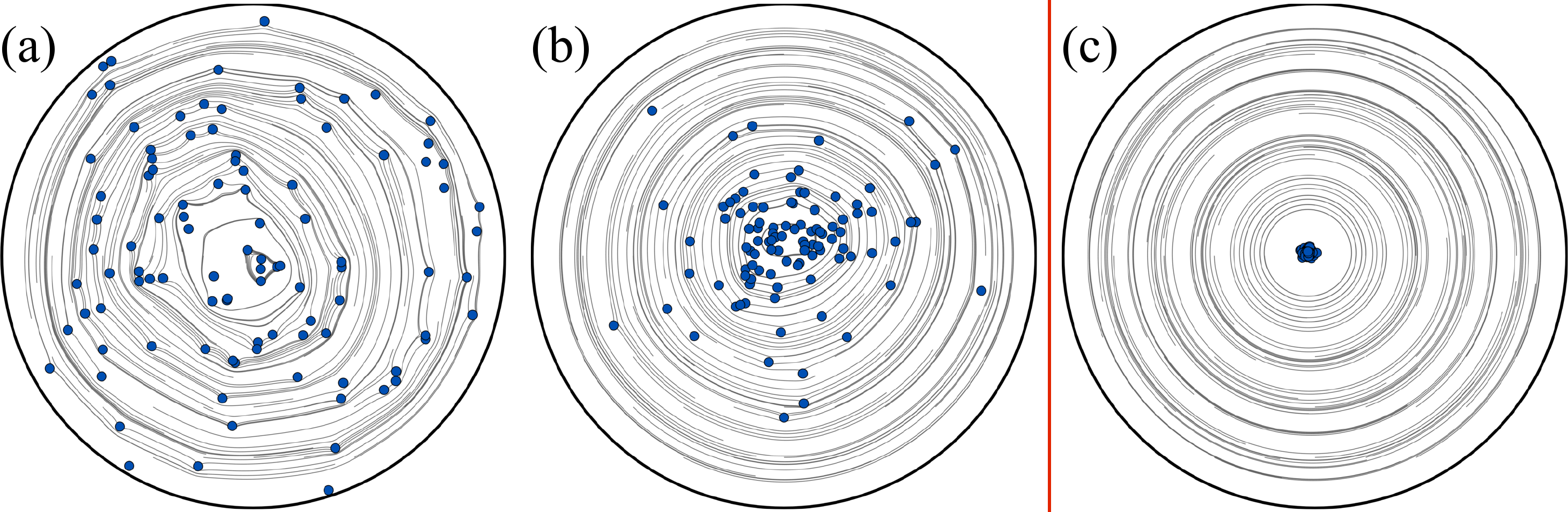}
\caption{
   Representative charge-polarised vortex configurations (a)-(c) at respective temperatures $T/T_{\rm EBC}= 2.000,1.031,$ and $0.769$, with $T_{\rm EBC}=-0.5\alpha N_v$ and $N_v=100$. The streamlines illustrate the velocity field generated by the collection of vortices. The red vertical line indicates the location of the condensation transition point between (b) and (c).}
\label{fig44}
\end{figure*}

Clustering of vortices and their condensation are two separate phenomena. Vortex clusters exist at all negative temperatures \cite{Groszek2017a}, where as non-zero condensate fraction only exist in the temperature range $ 0 > T > T_{\rm EBC}$. To demonstrate this more clearly, we have performed Monte Carlo calculations for a charge-polarised, $\sum_{i=1}^{N_v}s_i=N_v$, case where only one species of vortices is present in the system. Figure~\ref{fig44} (a)-(c) shows the vortex configurations at three different temperatures. These vortex configurations illustrate the fact that the vortex positions suddenly collapse when the radius of the host Onsager vortex cluster drops below a critical value, $R_c$. The transition illustrated in Fig.~\ref{fig1} corresponds to independent condensation of two species of vortices at the same temperature due to the equal numbers of vortices and antivortices. In vortex number imbalanced systems there are two, vortex number dependent, critical temperatures $T_{\rm maj} = -\alpha N_{\rm maj}/2$ and  $T_{\rm min} \approx -\alpha N_{\rm min}/2$.

The critical temperature for the condensation of an Onsager vortex in a single vortex species system may be predicted by a similar free energy argument as for two vortex species systems \cite{Kraichnan1980a,Simula2014a}. The Helmholtz free energy, $F=E-TS$, of a vortex configuration where all $N_v$ vortices are concentrated inside a circular region of radius $R^*$ is
\begin{equation}
F\approx \frac{\rho_s\kappa^2}{4\pi}N^2_v\ln\left(\frac{R}{R^*}\right) -T k_{\rm B}\ln\left(\frac{R^*}{R}\right)^{2N_v}
\label{FreeE}
\end{equation} 
where the energy $E$ is that of a multiply quantised vortex of core radius $R^*$ and the entropy $S$ is obtained as the logarithm of a statistical weight of the configuration. A change in the sign of the free energy signifies that the probability $p_{F}\propto e^{-F/k_{\rm B}T}$ of observing such a configuration becomes exceedingly likely and predicts a critical temperature
 \begin{equation}
 T_{\rm EBC}= -\frac{\alpha N_{v}}{2}= -\frac{\alpha N_{\rm tot}}{4},  
 \end{equation}
which is the same for a single species system with $N_v$ vortices as it is for a two-species system with the same number, $N_v= N_{\rm tot} /2 $, of vortices of one species. 

In the general imbalanced case with $N_+$ vortices and $N_-$ antivortices with $N_{\rm tot}=N_++N_-=N_{\rm maj}+N_{\rm min}$ and $N_{\rm maj}>N_{\rm min}$ there are two critical temperatures corresponding to separate condensation of each of the two vortex species. When the temperature approaches negative zero, the majority species condenses first at $T_{\rm maj} = -\alpha N_{\rm maj}/2$, followed by the condensation of the minority species at $T_{\rm min} \approx -\alpha N_{\rm min}/2$, where the latter is shifted slightly toward negative zero due to the interaction with the condensate of the majority species. 

The condensation of Onsager vortices may be viewed from the point of view of competition between solid body rotation within the core of the vortex cluster and potential flow outside the cluster, see Eqns~(\ref{velos}). Balancing the kinetic energy contributions of these two velocity fields in the mean-field interaction energy term in Eq.~(\ref{heff}) predicts  a critical cluster radius
\begin{equation}
R_c = e^{-1/4}R_\circ \approx 0.778800R_\circ,
\label{Rcrit}
\end{equation} 
such that for $T/T_{\rm EBC}>1$ the whole system prefers to mimic solid body rotation of a classical fluid, Fig.~\ref{fig44}(a) and (b), whereas for $T/T_{\rm EBC}<1$ the system prefers to mimic the velocity field of a quantised superfluid vortex, Fig.~\ref{fig44}(c). 

\begin{figure}[!h]
\centering
\includegraphics[width=0.8\columnwidth]{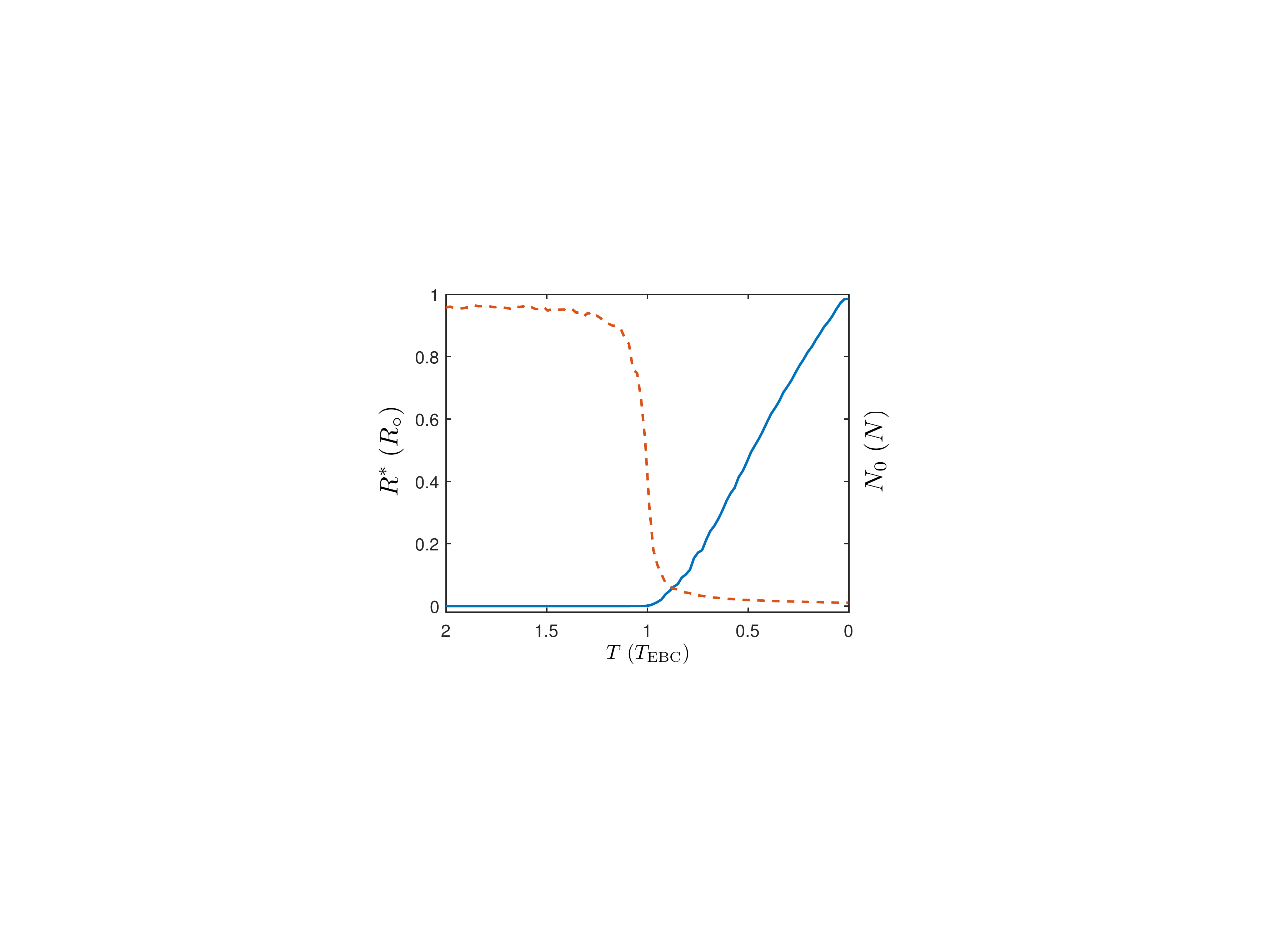}
\caption{
   Condensate fraction of $N_v=100$ single species, $s=+1$, vortices as a function of temperature (right axis, solid line) and the radius, $R^*$, of the vortex cluster, (left axis, dashed line), as a function of reduced temperature. 
}\label{fig55}
\end{figure}

Figure~\ref{fig55} shows the condensate fraction measured using the Eq.~(\ref{EBCfraction}). For $T/T_{\rm EBC}>1$, vortices are found scattered everywhere within the circular boundary and the condensate fraction is strictly zero. Near the transition, the vortices begin to clump and at critical radius $R_c$ the vortex cluster suddenly begins to collapse. Accompanied with the shrinking of the vortex cluster, the condensate fraction grows almost linearly with the reduced temperature. 

\begin{figure}[!t]
\centering
\includegraphics[width=1\columnwidth]{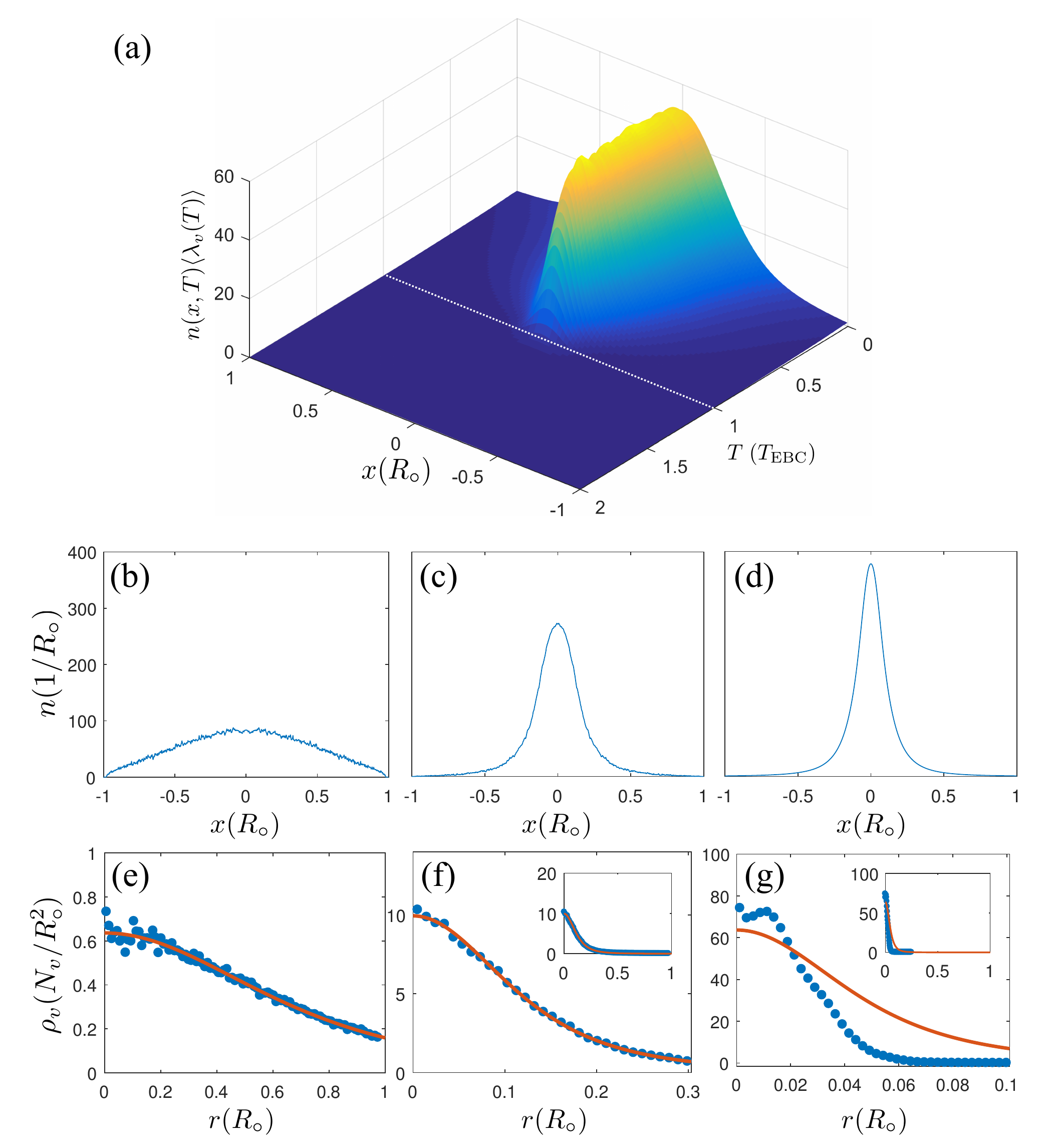}
\caption{
    Phase space density (a) as functions of position and reduced temperature, 1D vortex-particle densities (b)-(d) as functions of position corresponding to the representative charge-polarised vortex configurations at respective temperatures $T/T_{\rm EBC}= 2.000,1.031,$ and $0.769$, for $N_v=100$ with $T_{\rm EBC}=-0.5\alpha N_v$, and (e)-(g) the 2D density of the vortices in real space as functions of radial distance from the centre of the disc at temperatures corresponding to (b)-(d). The orange solid lines in (e)-(g) are least squares fits to theoretical vortex density function in Eq.~(\ref{vdensi}) with $T_{\rm fit}/T_{\rm EBC}= 2.004,1.033,$ and $(1.005)$. 
}\label{fig66}
\end{figure}

According to Eq.~(\ref{FreeE}) the specific heat at the transition
\begin{equation}
c_v = N^2_v\frac{\rho_s\kappa^2}{4\pi} \frac{1}{R^*} \frac{\partial R^*(T)}{\partial T}\Bigg|_{T_{\rm EBC}} \propto (T-T_{\rm EBC} )^{-1},
\label{spec heat}
\end{equation}
where in the last step we have assumed linear dependence of the cluster radius on the temperature in the vicinity of the transition, as suggested by Fig.~\ref{fig55}. 

Figure~\ref{fig66} (a) shows the phase space density, $n_v\langle\lambda_v\rangle$ of the vortices as functions of position and reduced temperature. The one-dimensional vortex-particle density $n(x)$ is obtained by modeling each vortex-particle by a normalised Gaussian wave packet of waist $\lambda_v$. The frames (b) - (d) show the 1D density $n(x)$ of the vortex gas for three different temperatures $T/T_{\rm EBC}= 2,1.031,$ and $0.769$. For $T/T_{\rm EBC}>1$ the vortex density is spread over the whole system while below the transition the vortex density becomes localised both in real space and in vortex momentum space. 

On approaching the condensation transition from the infinite temperature side the asymptotic form of the 2D real-space vortex density is predicted to be \cite{Lundgren1977a,Oneill1992a}
\begin{equation}
\rho_v(r) = \frac{1}{\pi} \left[{\left(1-\frac{T_{\rm EBC} }{T}\right)\left(1+\frac{T_{\rm EBC} }{T-T_{\rm EBC}}  r^2\right) ^2}\right]^{-1}
\label{vdensi}
\end{equation}
with normalisation 
\begin{equation}
 2\pi \int_0^{R_\circ} \rho_v(r) rdr =N_v.
\end{equation}

Figure~\ref{fig66} (e)-(g) shows the least squares fits of the function (\ref{vdensi}) to the radial 2D real space vortex density measured from the Monte Carlo calculations. Using the temperature $T$ as the sole fitting parameter the best fitting temperatures are measured to be $T_{\rm fit}/T_{\rm EBC}= 2.004,1.033,$ and $(1.005)$ and the resulting density profiles predicted by Eq.~(\ref{vdensi}) are shown as orange curves. The parentheses are used here to denote that the Eq.~(\ref{vdensi}) is used in a regime outside its validity. While the theory prediction, Eq.~(\ref{vdensi}), for the radial 2D vortex densities are in excellent agreement with the Monte Carlo data shown in Fig.~\ref{fig66} (e) and (f), the theory prediction in Fig.~\ref{fig66} (g) is clearly unphysical because Eq.~(\ref{vdensi}) diverges at $T=T_{\rm EBC}$ and cannot be used for predicting the vortex density in the condensed phase for which $T/T_{\rm EBC}<1$. This is evident in Fig.~\ref{fig66} (g) where the best fitting function has $T_{\rm fit}/T_{\rm EBC} =(1.005)$, as opposed to the actual temperature $T/T_{\rm EBC} =0.769$, of the state.

\begin{figure}[!t]
\centering
\includegraphics[width=0.8\columnwidth]{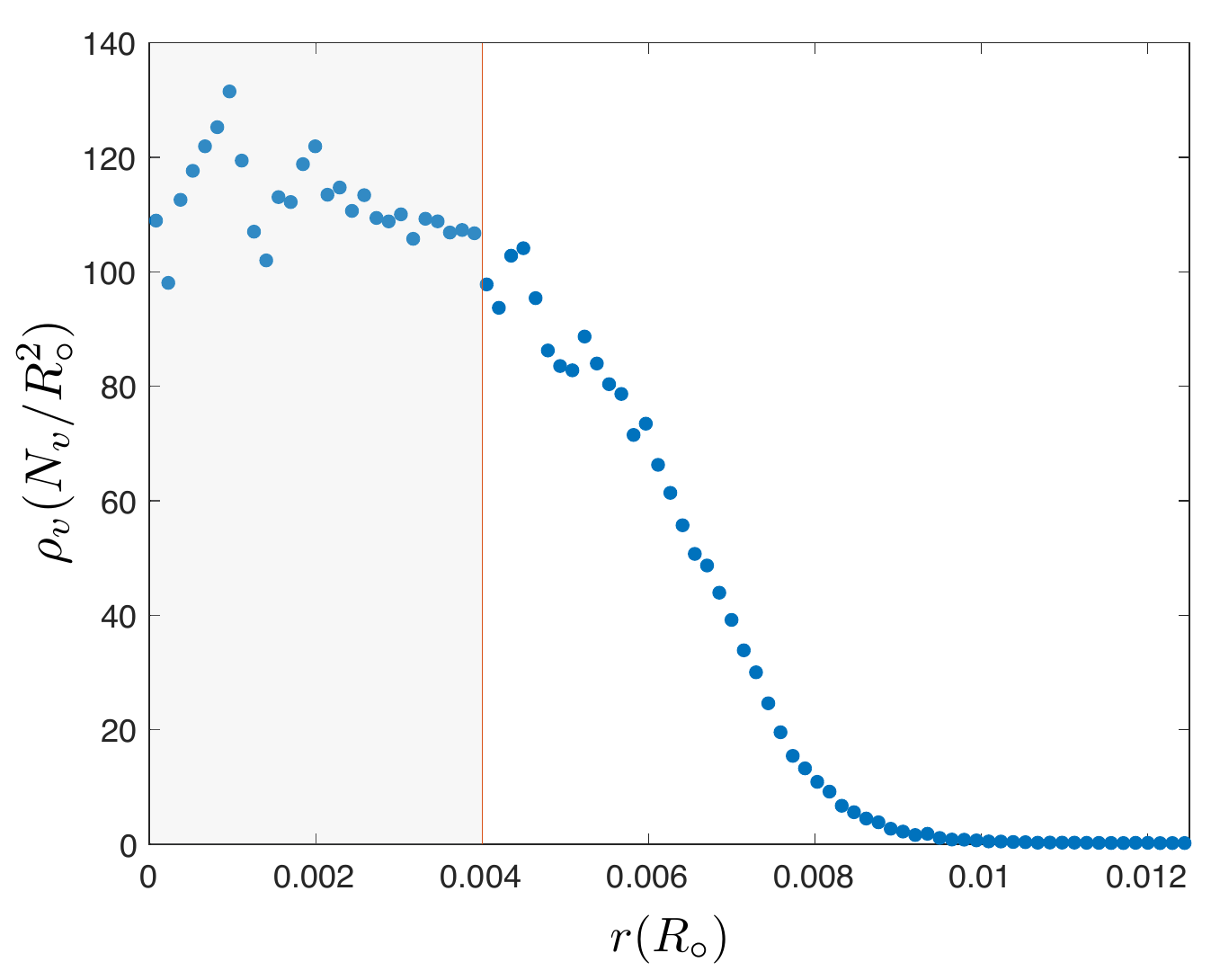}
\caption{
   Two-dimensional vortex density in real space as a function of radial distance measured from the centre-of-mass of the vortex cluster. The temperature, condensate fraction and condensate radius of this state are, respectively, $T/T_{\rm EBC}= 0.244$, $N_0/N=0.773$, and $R_{\rm cond} /R_\circ = 0.004$. The condensate corresponds to the shaded area with flat, incompressible, density of the vortex fluid. 
}\label{fig10}
\end{figure}

Figure~\ref{fig10} shows the two-dimensional vortex density as a function of radial position for  $T/T_{\rm EBC}=0.244$. The condensed vortices seem to form a fluid like incompressible core of the cluster with a constant vortex density. The number of vortices within the shaded region corresponding to the condensate is equal to $N_0$. 

\section{Discussion}

The physics of the vortex system in the vicinity of the critical negative absolute temperature 
\begin{equation}
T_{\rm EBC} = - \frac{\hbar \omega_k}{2 k_{\rm B}}\frac{m_0}{m} N_{v} = - \frac{\hbar \omega_k}{4 k_{\rm B}}\frac{m_0}{m} N_{\rm tot},
\end{equation}
where $\omega_k$ is the kelvon frequency and $m_0$ is the vortex mass per unit length \cite{Simula2017a}, has been discussed extensively in the recent literature \cite{White2012a,Reeves2013a,Billam2014a,Simula2014a,Reeves2015a,Stagg2015a,Groszek2016a,Skaugen2016a,Yu2016a,Skaugen2016b,Salman2016a,Groszek2017a,Esler2017a} yet the nature of the condensate has remained unclear. This is partly because of the divergent behaviour of the zero-core point vortex model that becomes invalid at the critical point of condensation and is unable to yield predictions for the condensed phase. The situation is the same as in the positive temperature side where the Hauge--Hemmer transition to pair-collapsed phase in the zero-core point vortex model is divergent and the structure of the vortex core, which is always present in any real physical system, must be accounted for. Including the effects of non-zero vortex cores in the positive temperature systems allows correct treatment of the Berezinskii--Kosterlitz--Thouless phase transition whose critical temperature is shifted by a factor of 2 with respect to the Hauge--Hemmer transition that occurs at $T_{\rm HH}=\alpha$ \cite{Hauge1971a}. Similarly, any self-consistent treatment of the negative absolute temperature Onsager vortex condensate must include the effects of non-zero size of the vortex core. For the sake of clarity, we discuss the one and two vortex species cases separately below.

\subsection{One vortex species case} 
First, considering the single vortex species system shows that the condensation of the Onsager vortices occurs when the vortex cores within a cluster of vortices begin to merge into a single vortex structure with multiple circulation quanta, signifying the emergence of large degeneracy in the quasiparticle degrees of freedom of the vortices.  

We briefly recall the underpinnings of the quantum Hall effect of two-dimensional electron gas in a strong external magnetic field corresponding to extremely large kinetic energy per electron. This 2D problem is often theoretically mapped onto a dual 1D harmonic oscillator problem, which reveals that the topological phase transitions to the integer quantum Hall states occur when the electrons condense in the highly degenerate lowest Landau level. Although the electrons move in 2D space, the topological phase transitions are quantified in terms of the eigenstates of a 1D harmonic oscillator.

Similar physics is pertinent to the Onsager vortex condensation transition. It is therefore useful to consider the closest known physical realisation of the Onsager's point vortex model, which is a Bose--Einstein condensate with quantised vortices nucleated in the macroscopic condensate wavefunction. A trial wave function for such a system may be expressed as
\begin{equation}
\Psi({\bf r}) = \prod_{j=1}^{N_v} \chi_j(r) \sqrt{f}e^{i\theta_j},
\label{and}
\end{equation}
where $f$ is the smooth condensate particle density in the absence of the vortices, $\theta_j = \arg[(x-x_j) +i(y-y_j)]$ are the additive phase functions with singularities at the vortex locations $\{x_j,y_j\}$, and the (soft) vortex core function
\begin{equation}
\chi_j(r) = \frac{r^2}{\sqrt{r^2 +2\xi^2}}.
\end{equation}

The probability current 
\begin{equation}
{\bf j} = \frac{\hbar}{2mi}[\Psi({\bf r})^* \nabla \Psi({\bf r})  - \Psi({\bf r}) \nabla \Psi^*({\bf r}) ] \equiv |\Psi({\bf r})|^2 {\bf v}_s
\end{equation}
of the state (\ref{and}) defines the superfluid velocity field ${\bf v}_s$, the incompressible component of which is remarkably well approximated by the velocity field of Onsager's point vortex model \cite{Groszek2017b}.

The elementary excitation spectrum of a 2D vortex configuration is obtained by solving the Bogoliubov--de Gennes (BdG) eigenvalue problem \cite{Fetter1972a}. The $N_v$ phase singularities due to the $N_v$ quantised vortices in the system yield $N_v$ low energy quasiparticle eigenstates \cite{Simula2013a} that satisfy the bosonic commutation relations
\begin{equation}
[\eta_q,\eta_p^\dagger]=\delta_{q,p} ; \hspace{5mm}[\eta_q,\eta_p]=[\eta^\dagger_q,\eta^\dagger_p]=0, 
\end{equation} 
where $\eta_q^\dagger$ and $\eta_q$ are the usual Bogoliubov quasiparticle particle creation and annihilation operators. In accordance with the quasiparticle picture of superfluids, the macroscopic multiply connected wave function, Eq.~(\ref{and}), may be expressed in terms of the countably infinite set of such quasiparticle states \cite{Fetter1972a}. These Bogoliubov quasiparticles are bosons and this property is inherited by the host vortices whose circulation is quantised. 

For a single vortex with $N_v$ circulation quanta the condensate wave function may be expressed as 
\begin{equation}
\Psi({\bf r}) =  \chi_{N_v}(r) \sqrt{f}e^{iN_v \theta},
\label{multiqv}
\end{equation}
where $\chi_{N_v}(r)$ is the structure function of the vortex core with $N_v$ circulation quanta. The BdG quasiparticle excitation spectrum of such a state has only one vortex eigenmode, corresponding to the one phase singularity, with orbital angular momentum quantum number equal to $w=-N_v$ \cite{Simula2002a}. This high-winding number bosonic quasiparticle mode is a BEC of $N_v$ Bogoliubov quasiparticles associated with the $N_v$ vortex circulation quanta, in essence forming a ``vortex BEC in a BEC of atoms". Such quasiparticle condensates are not unusual. For example, magnons (spin-waves) have previously been observed to form Bose--Einstein condensates of their own within their host BECs \cite{Volovik2008a,Demokritov2006a,Vainio2015a}.

The circulation 
\begin{equation}
\Gamma = \oint _C{\bf v}\cdot d {\boldsymbol \ell} 
\end{equation}
of a classical point vortex measured along a path $C$ that encloses the vortex is invariant with respect to continuous deformations of the path $C$ precisely as for a  quantised vortex in a BEC of atoms. In a BEC of atoms the vortex cores trap the bosonic quasiparticles and when these localised bosonic modes overlap they may form a condensate. The vortex density of the point-like vortex cores thus effectively measures the density of states of the Bogoliubov quasiparticles attached to the vortices and the overlapping of the vortex cores is tantamount to the condensation of the $N_v$ BdG quasiparticles associated with the vortex degrees of freedom. It is in this sense that the classical point-vortex Hamiltonian describes the bosonic degrees of freedom of the quantised vortices and their quantum statistical condensation at $T_{\rm EBC}$. Indeed, Eq.~(\ref{FreeE}) applies equally well for both classical point vortices and for quantised vortices in a BEC.

The vortex-particle duality allows a 1D treatment of the 2D vortex gas and motivates the definition of the condensate fraction as
\begin{equation}
\frac{N_0}{N} =\frac{\xi^2}{\langle r_{\rm nn} \rangle^2}. 
\label{EBCfractionone}
\end{equation}
The condensate fraction is equal to the area ratio of the minimum possible phase-space area occupied by the $N_v$ vortices to the area actually occupied by them. A high vortex condensate fraction is equivalent to strong overlap between the BdG quasiparticle modes of the quantised vortices. The point-vortex model description works well in this extreme states of vortices because in such situations kinetic energy of the BEC of atoms is overwhelmingly larger than the usual mean-field atom-atom interaction.

It is interesting to recall the structure of a simple vortex in a superfluid or a superconductor. Outside the vortex core the superfluid or superconducting order parameter is at its bulk value whereas in the vortex core region the superfluid order parameter vanishes and the original symmetry of the full Hamiltonian is locally restored. A local observer spatially traversing across a vortex core in such systems measures superfluid-normal-superfluid phase changes along their path.

For a 2 vortex problem, the change in the phase space topology of the point vortex model has been quantified by identification of the phase-space wall that divides the two regions of phase space where the vortex trajectories are either overlapping or non-overlapping \cite{Murray2016a}. We conjecture that similar phase-space dividing wall is associated with any number $N_v$ of vortices and that associated with the condensation of Onsager vortices a multiply connected phase-space topology transforms to a single connected region. In this sense, the condensation of Onsager vortices should be viewed as a topological phase transition.

\subsection{Two vortex species case} 
In the neutral two vortex species case, the Onsager vortex condensation transition described above for single vortex species systems occurs in both of the vortex types separately, and simultaneously. In the case of vortex-imbalanced system the majority vortex species condenses first at
\begin{equation} 
 T_{\rm maj} = -\alpha N_{\rm maj}/2,
\end{equation}
followed by the condensation of the minority species at
\begin{equation} 
 T_{\rm min} \approx -\alpha N_{\rm min}/2,
\end{equation}
where $T_{\rm min}$ shifted slightly toward negative zero due to the interaction with the condensate of the majority species. 

Before the condensation of the Onsager vortices proceeds, the vortices become spatially phase separated as shown e.g. in Fig.~\ref{fig1} and in the Supplemental Figure~S1(b) of Ref.~\cite{Simula2014a}. Such phase separation is one step in the sequence of phase space compactification leading to the EBC transition. The critical temperature is the same for a single species system with $N_v$ vortices as it is for a two-species system with the same number, $N_v= N_{\rm tot} /2 $, of vortices of one species, and the condensation transition occurs in both systems. In contrast, the phase separation is specific to the two-species case.

\section{Conclusions}

In conclusion, we have employed a vortex--particle duality to establish a correspondence between vortices in a two-dimensional fluid and a one-dimensional gas of vortex particles. Using this mapping, we have provided a quantitative measure for the condensation of Onsager vortices. The vortex condensate forms due to the overlap of the vortex cores. Ultimately, deep in the condensed phase a phase-space Wigner crystallisation of vortices with hard cores takes place \cite{Simula2014a} while soft core vortices would yield multiple quantum vortex state \cite{Simula2002a}. The situation bears resemblance to rapidly rotating neutral superfluids that are predicted to undergo phase changes when the vortex cores begin to significantly overlap and the filling factor, or the number of fluid particles per vortex, approaches unity \cite{Cooper2008a,Fetter2009a}. One interesting future direction will be to study connections between the 1D vortex particle theory and other 1D systems \cite{Tercas2013a,Tercas2014a,Giamarchi2003a}. 

Although observation of negative absolute temperatures is most transparent in neutral vortex gas where absolute negative temperature states are readily associated with the emergence of conspicuous vortex clusters \cite{Gauthier2018a,Johnstone2018a}, it seems that the most suitable system to study the critical physics of Einstein--Bose condensation of Onsager vortices is a single species vortex system. We therefore propose an experiment to observe the condensation of Onsager vortices using a BEC or superfluid Fermi gas of atoms by creating a giant vortex with multiple circulation quanta using, e.g., topological phase imprinting \cite{Leanhardt2002a} or high-winding number Laguerre--Gauss laser beams \cite{Andersen2006a}, to imprint a multiply quantised, $w=N_v$, quantum vortex into a superfluid in a preferably uniform trap \cite{Henderson2009a,Gaunt2013a,Chomaz2015a,Bell2016a,Lee2015a,Gauthier2016a,Mukherjee2017a,Hueck2018a,Johnstone2018a,Gauthier2018a}. Subsequently monitoring the decay of the state into $N_v$ singly quantised vortices, evolving from configurations akin to Fig.~\ref{fig44}~(c) to those shown in (b), will enable quantitative observation of crossing the critical temperature $T_{\rm EBC}$. Additional benefit of this approach is that it does not require detection of the vortex circulation signs. Direct measurement of the vortex positions and their core sizes enables direct measurement of the condensate fraction $N_0/N$, Eq.~(\ref{EBCfraction}), in the condensed phase for $T/T_{\rm EBC}<1$. Equation~(\ref{vdensi}) enables explicit and accurate measurement of the vortex temperature for $T/T_{\rm EBC}>1$, as shown in Fig.~\ref{fig66} (e) and (f). In combination, these two measurements will enable direct and quantitative experimental observation of the condensation transition of the Onsager vortices.  


\section*{Acknowledgments}
We acknowledge support from an Australian Postgraduate Award (R.N., A.G.), the Australian Research Council via Discovery Project No. DP130102321 and DP170104180 (T.S.), and the nVidia Hardware Grant Program. This research was undertaken with the assistance of resources from the National Computational Infrastructure (NCI), which is supported by the Australian Government.


%

\end{document}